\documentclass[a4paper,12pt]{article}
\usepackage{epsfig}
\usepackage{amssymb}
\usepackage{amsfonts}

\newskip\humongous \humongous=0pt plus 1000pt minus 1000pt

\newif\ifdtup

\catcode`\@=11

\@addtoreset{equation}{section}
\def\theequation{\thesection.\arabic{equation}}

\def\@normalsize{\@setsize\normalsize{15pt}\xiipt\@xiipt
\abovedisplayskip 14pt plus3pt minus3pt%
\belowdisplayskip \abovedisplayskip
\abovedisplayshortskip \z@ plus3pt%
\belowdisplayshortskip 7pt plus3.5pt minus0pt}

\def\small{\@setsize\small{13.6pt}\xipt\@xipt
\abovedisplayskip 13pt plus3pt minus3pt%
\belowdisplayskip \abovedisplayskip
\abovedisplayshortskip \z@ plus3pt%
\belowdisplayshortskip 7pt plus3.5pt minus0pt
\def\@listi{\parsep 4.5pt plus 2pt minus 1pt
     \itemsep \parsep
     \topsep 9pt plus 3pt minus 3pt}}

\relax

\catcode`@=12

\catcode`\@=11

\def\section{\@startsection{section}{1}{\z@}{3.5ex plus 1ex minus
   .2ex}{2.3ex plus .2ex}{\large\bf}}

\def\thesection{\arabic{section}}
\def\thesubsection{\arabic{section}.\arabic{subsection}}

\def\appendix{\setcounter{section}{0}
 \def\thesection{Appendix \Alph{section}}
 \def\thesubsection{\Alph{section}.\arabic{subsection}}
 \def\theequation{\Alph{section}.\arabic{equation}}}

\begin{document}

%%%%%%Define some new commands and  macros
\newcommand{\beq}{\begin{equation}}
\newcommand{\eeq}{\end{equation}}
\newcommand{\bea}{\begin{eqnarray}}
\newcommand{\eea}{\end{eqnarray}}
\newcommand{\beas}{\begin{eqnarray*}}
\newcommand{\eeas}{\end{eqnarray*}}
\newcommand{\defi}{\stackrel{\rm def}{=}}
\newcommand{\non}{\nonumber}
\newcommand{\bquo}{\begin{quote}}
\newcommand{\enqu}{\end{quote}}
\newcommand{\mat}{\mathbf}
%%%%%%%%%%%%%%%%%%%%%%%%%%%%%%%%%% definitions
\def\de{\partial}
\def\Tr{ \hbox{\rm Tr}}
\def\const{\hbox {\rm const.}}
\def\o{\over}
\def\im{\hbox{\rm Im}}
\def\re{\hbox{\rm Re}}
\def\bra{\langle}\def\ket{\rangle}
\def\Arg{\hbox {\rm Arg}}
\def\Re{\hbox {\rm Re}}
\def\Im{\hbox {\rm Im}}
\def\diag{\hbox{\rm diag}}
\def\longvert{{\rule[-2mm]{0.1mm}{7mm}}\,}
\def\Z{\mathbb Z}
\def\N{{\cal N}}
\def\tD{{\widetilde  D}}
\def\tq{{\widetilde q}}
\def\W{{\cal W}}
\def\tQ{{\widetilde Q}}
\def\dag{{}^{\dagger}}
\def\p{{}^{\,\prime}}
\def\a{\alpha}
\def\Tr{ \hbox{\rm Tr}}
\def\tM{{\widetilde M}}
\def\tm{{\widetilde m}}
\def\T{{\cal T}}
\def\t{T}
\def\J{{\cal J}}
\def\tk{{\widetilde k}}
\def\tz{{\widetilde z}}
\def\td{{\widetilde d}}

\begin{titlepage}
\begin{flushright}
hep-th/0412241\\
\end{flushright}

\bigskip

\begin{center}
{\Large

{\bf  
The Holomorphic Tension of Nonabelian Vortices
}

 } 
\end{center}

\renewcommand{\thefootnote}{\fnsymbol{footnote}}
\bigskip
\begin{center}
{\large   Stefano Bolognesi }
 \vskip 0.20cm
\end{center}

\begin{center}
{\it      \footnotesize
Scuola Normale Superiore - Pisa, Piazza dei Cavalieri 7, Pisa, Italy \\
\vskip 0.10cm
and\\
\vskip 0.10cm
Istituto Nazionale di Fisica Nucleare -- Sezione di Pisa, \\
Via Buonarroti 2, Ed. C, 56127 Pisa,  Italy   \\  }
\vskip 0.15cm
s.bolognesi@sns.it\\
\end {center}

\setcounter{footnote}{0}

\bigskip
\bigskip

\noindent  
\begin{center} {\bf Abstract} \end{center}
We continue the work hep-th/0411075 considering here the case of degenerate masses. A nonabelian vortex  arises  in $r$-vacua upon the breaking by a superpotential for the adjoint field. We find the BPS tension in the strong coupling regime computing the dual-quark condensate. Then we find that it is equal to a simple quantity in the chiral ring of the theory and so we conjecture the validity of our result out of the strong coupling regime. Our result gives an interesting hint about the duality $r \leftrightarrow N_f-r$, seeing it as the exchange first $\leftrightarrow$ second sheet of $\N=1$ Riemann surface.

\vfill

\begin{flushleft}
December, 2004
\end{flushleft}

\end{titlepage}

\bigskip

\hfill{}

\section{Introduction}

In \cite{Stefano} we studied the tension of vortices in ${\cal N}=2$ SQCD (in the case of non degenerate masses) broken to ${\cal N}=1$ by a superpotential $W(\Phi)$, in color-flavor locked vacua. When there is a color-flavor locking some flavors become massless in the $\N=2$ theory.  Due to the presence of the superpotential, these flavors condense and create a vortex solution. The tension of the vortex can be written as a BPS tension plus a non-BPS contribution
\beq
\label{claim}
T=T_{BPS} + T_{non\, BPS}\ ,
\eeq
where 
\beq
T_{BPS}=4\pi|\T| 
\eeq 
 and we called $\T$ the {\it  holomorphic tension}.  This is directly related to the central charge of the supersymmetry algebra. If Lorentz invariance is broken by a vortex configuration, one can introduce a central charge in the $\N=1$ superalgebra. This central charge is essentially the holomorphic tension $\T$.

Here we consider the case of $N_f$ flavors with degenerate mass $m$. There are vacua in the moduli space of the $\N=2$ theory, where a non abelian subgroup $SU(r) \times U(1)^{N_c-r+1}$ survives in the infrared \cite{APS,CKM}. In the weak coupling regime ($m>>\Lambda$) $N_f$ quarks are locked to $r$ colors and if $r <  N_f/2$ the nonabelian $SU(r)$ is infrared free and survives at low energy.   The breaking to $\N=1$ by a soft mass for the adjoint field and the formation of the nonabelian vortex  have been studied in \cite{vortici} in the weak coupling regime (similar solutions where also found in \cite{HT}). The condensation of quarks creates a nonabelian vortex that confines the massive nonabelian magnetic monopoles.  For general value of $m$ the low energy degrees of freedom are no more the quarks $Q, \tQ$ but the  dual-quarks $\tD, D$.
Here we compute in the strong coupling regime the dual-quark condensate using the factorization equations \cite{CDSW,largeN,factor} that relate  the $\N=2$ Seiberg-Witten curve  to the $\N=1$ matrix model curve. The last one is defined by 
\beq
\label{curva1intr}
 {y_m}^2={W\p}^2(z)+f(z)\ ,
\eeq 
 where $f(z)$ is a polynomial (a quantum correction) that depends on the vacuum.
The result for the dual-quark condensate is
\bea
\tD_I D_J=
\left(
\begin{array}{cccc}
-\left.\sqrt{{W\p}^2+f}\right|_{z=m} {\bf 1}_{{\rm min}(r,N_f-r)}&&&\\
&0 &&\\
&& \ddots &\\
&&& 0 \\
\end{array}\right) \ ,
\eea
where $I,J=1 \dots N_f$ are flavor indices. 
This condensate creates a nonabelian vortex whose holomorphic tension is
\beq
\label{result}
\T = \left.\sqrt{{W\p}^2+f}\right|_{z=m} \ .
\eeq

In the case of $r$-vacua we can say something more with respect to \cite{Stefano}. We can express the dual-quark condensate as a simple quantity in the chiral ring of the theory.  Thanks to a generalized version of the Konishi anomaly \cite{CDSW,CSW} one can compute all the expectation values of the operators in the chiral ring of the theory. The generators of the chiral ring are 
\beq
\label{convenctionintr}
T(z)=\Tr\, \frac{1}{z-\Phi}\ ,\qquad
R(z)=-\frac{1}{16\sqrt{2}\pi^2}\Tr\,\frac{W^{\a}W_{\a}}{z-\Phi} \ ,
\eeq
\beq
M_I(z)=\tQ_I \frac{1}{z-\Phi}Q_I \ .
\eeq
 These are differential forms on the Riemann surface (\ref{curva1intr})
with punctures at $q$ and $\tq$ (two points with the same $z=m$ coordinate and respectively in the first and in the second sheet of the Riemann surface).   
The dual-quark condensate is exactly equal to the residue 
\beq
\label{aaaafirst}
\tD_{I} D_{I}= \frac{1}{2\pi i} \oint_m  M_{I}(z) dz   \ ,
\eeq
evaluated  around $q$ if the number of locked quarks $r$ is less than $N_f/2$ or  around  $\tq$ if $r$ is greater than $N_f/2$.
Note that the condensate $\tD_{I} D_{I}$ has no meaning out of the strong coupling regime. The residue $\frac{1}{2\pi i} \oint  M_{I}(z) dz$ is instead computed using the generalized Konishi anomaly and so it is not restricted to the strong coupling regime.

The paper is organized as follows. In Section \ref{Model} we introduce the model and we compute the tension in the classical limit. In Section \ref{holo} we compute the holomorphic tension in the strong coupling regime. In Section \ref{chiral} we express the dual-quark condensate as a simple object in the chiral ring of the theory.  \ref{NAV} is devoted to the explicit construction of the nonabelian vortex and its moduli space.

\section{The Model \label{Model}}

Our model is $\N=2$ $U(N_c)$ gauge theory with $N_f$ flavors of mass $m$, broken to $\N=1$ by a superpotential for the adjoint field.
The Lagrangian is
\bea
\label{QFT}
{\cal L}&=&\int d^2\theta \,\frac{1}{2g^2}\Tr_{N_c} \, (W^{\alpha}W_{\alpha}) +h.c. \\ &&+ \int d^2\theta d^2\bar{\theta}  \,\frac{2}{g^2}\Tr_{N_c} \, (\Phi\dag e^V\Phi e^{-V})+ \int d^2\theta d^2\bar{\theta} \,\sum_{I=1}^{N_f} (Q_I\dag e^{V}Q_I+\widetilde{Q}_I e^{-V}\widetilde{Q}^{\dagger }_I )\nonumber\\
&& + \int d^2\theta \,\W + h.c.  \ , \nonumber
\eea
where the superpotential is
\beq
\W= \sum_{I=1}^{N_f}  \sqrt{2} (\widetilde{Q}_I\Phi Q_I - m \widetilde{Q}_I Q_I )+\sqrt{2} \Tr_{N_c} W(\Phi) \ ,
\eeq
and
\beq
W(z)=\sum_{j=0}^{k} \, \frac{g_j}{j+1}z^{j+1}\ ,\qquad W\p(z)=g_k\prod_{j=1}^{k}\, (z-a_j)\ .
\eeq
The flavor are labeled by the index $I=1,\dots,N_f$.

The $F_{Q_I}$ and $F_{\tQ_I}$ terms set to zero imply that the diagonal elements of $\phi$ are equal either to $m$ or to a root of $W\p$
\beq
\label{phicondensate}
\langle\phi\rangle=\left(\begin{array}{cccc}
m {\bf 1}_{r}&&&\\
&a_1 {\bf 1}_{N_1}&&\\
&&\ddots&\\
&&&a_n {\bf 1}_{N_n}\\
\end{array}\right)\ ,
\eeq
where
\beq
\sum_{j=1}^{n} \, N_j+r=N_c\ .
\eeq
$r$ is the numbers of flavors that are locked to $m$ and must be not greater than the number of flavors, $0 \leq r \leq N_{f}$.  
The gauge group is broken by (\ref{phicondensate}) to $ U(r) \times \prod_{j=1}^{n} U(N_j)$.

If $r$ colors and flavors are locked at the same eigenvalue,  then in the low energy we have also a massless hypermultiplet in the fundamental of $U(r)$.  The condensation of this hypermultiplet breaks the $U(r)$ group and creates a nonabelian vortex that confines nonabelian magnetic monopoles \cite{vortici}.   The $F_{\phi}$ term and the $D$ term together yield the potential for the quark fields
\beq
V= g^2 \Tr_{r} \, (|q\tq +W'|^2)+\frac{g^2}{4}\Tr_{r} \, ((qq\dag -\tq\dag \tq)^2)\ ,
\eeq
where we have suppressed the gauge and flavor indices, which are summed.
This may be expressed in a $SU(2)_R$ invariant form using the doublet $q^{\alpha}=(q,\tq\dag)$:
\beq
V=\frac{g^2}{2}\Tr_{r} \Tr_2 \, ({q}^{\alpha}{q\dag}_{\beta}-\frac{1}{2}{\delta^{\alpha}}_{\beta}{q}^{\gamma} {q\dag}_{\gamma}-\xi_a{(\sigma_a)^\alpha}_\beta)^2\ , 
\eeq
\[ -\xi_1+i\xi_2=W\p(m)\ ,\qquad \xi_3=0\ .
\]
An $SU(2)_R$ rotation brings the potential to a  form with a  Fayet-Iliopoulos (FI) term $-\int d^2\theta
d^2 \bar{\theta} \, 2v V $ and no superpotential
\beq
\label{potential}
V=\frac{g^2}{4} \Tr_{r} \, ((qq\dag -\tq\dag \tq -2v {\bf 1}_{r} )^2)\ ,
\eeq
where $v=|W\p(m)|$. The low energy Lagrangian is thus
\beq
{\cal L}\supset -\frac{1}{2g^2}\Tr_{r} \, (F_{\mu\nu}F^{\mu\nu}) - (D_\mu q)(D_\mu q)-\frac{g^2}{4}\Tr_{r}\,((qq\dag -2{v}{\bf 1}_{r})^2)\ .
\eeq
The nonabelian BPS vortex is explicitly build in \ref{NAV} where we consider also the possibility that the $SU(r)$ coupling is different from the $U(1)$ coupling. 
At the end we will find that the tension is $T_{BPS}=4\pi|W\p(m)|$ and so the holomorphic tension is 
\beq
\T=W\p(m) \ .
\eeq

\section{Strong Coupling \label{holo}}

\begin{figure}[ht]
\begin{center}
\leavevmode
\epsfxsize 12   cm
\epsffile{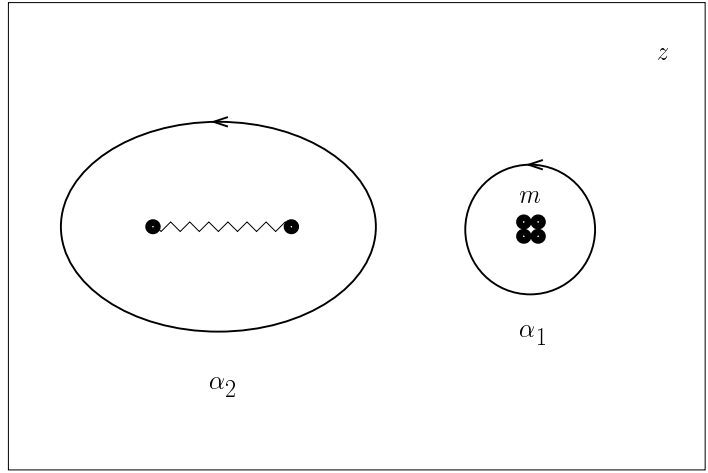}    
\end{center} 
\caption{Cycles in $U(3) \to SU(2) \times U(1) \times U(1)$ theory.}
\label{cicle} 
\end{figure}

Now we compute the holomorphic tension computing the dual-quark condensate as in \cite{Stefano}. In this section we will denote the quantum parameter 
\beq
{r_q}={\rm min}(r,N_f-r) \ .
\eeq
The simplest case to consider is $k=n=N_c-{r_q}$. To avoid some technicalities we compute the condensate only in this case.
The factorization of the  $\Sigma_{\N=2}$ curve (\cite{CDSW,factor}) gives 
\beq
\label{factorU(2)}
 \qquad y^2=\frac{1}{{g_{N_c - {r_q}}}^2}({W\p}^2+f)(z-m)^{2{r_q}} \ .
\eeq
See Figure \ref{cicle} for the case $N_c=3$ and ${r_q}=2$.  
In the low energy we have a $\N=2$ $SU({r_q}) \times U(1) \times U(1)^{N_c-{r_q}}$ gauge theory with hypermultiplet $\tD, D$ with charges given in Table \ref{low}. 
\begin{table}[ht]
\begin{center}
\begin{tabular}{ccccc}
& $U(N_f)$ & $SU({r_q})$ & $U(1)$ & $U(1)^{N_c-{r_q}}$ \\
\hline
&  $\underline{N}_f$&$\underline{{r_q}}$&$1$&$0$\\
\hline
\end{tabular}
\end{center}
\caption{Low energy in ${r_q}$-vacua.}
\label{low}
\end{table}
The low energy superpotential is
\beq
\label{lowsup}
\W_{low}=\sqrt{2} \sum_{I=1}^{N_f}\, ( \tD_I A_{\underline{{r_q}}} D_I +  \tD_I A_{1} D_I) + \sqrt{2} W_{eff}(A_{\underline{{r_q}}},A_1,\dots,A_{N_c-{r_q}+1}) )\ ,
\eeq
where the effective superpotential that breaks to $\N=1$ is
\beq
W_{eff}=\sum_{j=1}^{N_c-{r_q}} \, g_j u_{j+1}(A_{\underline{{r_q}}},A_1,\dots,A_{N_c-{r_q}+1}) \ .
\eeq
$A$'s are the chiral superfields of the $\N=2$ gauge multiplets, $A_{\underline{{r_q}}}$ is the one of the $SU({r_q})$ gauge multiplet, $A_1$ is the one of the $U(1)$ multiplet coupled to the dual quarks. To compute the tension we need to consider only the $F_A$ terms of the potential:
\beq
F_{A_{\underline{{r_q}}}}=2 {g_{na}}^2 \sum_{b=1}^{r_q} | \sum_{I=1}^{N_f} \td_I T^b d_I |^2 \ , 
\eeq
\beq
F_{A_1}=2 {g_1}^2 |\sum_{I=1}^{N_f} {\td_I} {d_I}   +\frac{\de W_{eff}}{\de a_1}|^2\ ,  
\eeq
\beq
F_{A_s}=2 {g_s}^2 |\frac{\de W_{eff}}{\de a_s}|^2\ , \qquad s=2,\dots,N_c-{r_q}+1 \ . 
\eeq
Making the BPS ansatz $\td=-d\dag$ we obtain the action for the nonabelian BPS vortex studied in \ref{NAV}. The holomorphic tension is given setting to zero the $F_{A_1}$ term
\beq
\label{T}
\T=-\frac{1}{{r_q}} \sum_{I=1}^{N_f} {\td_I}  {d_I} =  \frac{1}{{r_q}}\frac{\de W_{eff}}{ \de a_1} \ ,
\eeq
while the other $F_{A}$'s  terms give stationary conditions
\beq
\label{S}
0=\frac{\de W_{eff}}{ \de a_{s>1}}
\eeq
Writing (\ref{T}) and (\ref{S}) in a matrix form and multiplying by the inverse matrix we obtain:
\beq
\label{matricedue}
\left(
\begin{array}{c}
g_0\\
\vdots\\
\vdots\\
g_{N_c - {r_q}}\\
\end{array}
\right)
=
\left(
\begin{array}{cccc}
\de a_1/\de u_1&\dots&\dots&\de a_{N_c - {r_q}+1}/\de u_1\\
\vdots&&&\vdots\\
\vdots&&&\vdots\\
\de a_1/\de u_{N_c - {r_q}+1}&\dots&\dots&\de a_{N_c - {r_q}+1}/\de u_{N_c - {r_q}+1}\\
\end{array}
\right)
\left(
\begin{array}{c}
{r_q} \T\\
0\\
\vdots\\
0\\
\end{array}
\right)\ .
\eeq
The last equation is
\beq
g_{N_c-{r_q}}=\frac{\de a_1}{\de u_{N_c - {r_q}+1}} {r_q} \T \ ,
\eeq
where by the Seiberg-Witten solution (see \cite{Stefano} for more details\footnote{To reproduce the correct semiclassical result, the normalization of the holomorphic differential here must be $1/{r_q}$ the ones used in \cite{Stefano}. Here, in fact, we are making an integral around $2{r_q}$ collided roots while in \cite{Stefano} the integral was around $2$ collided roots.})
\beq
\frac{\de a_1}{\de u_{N_c - {r_q}+1}}=\frac{1}{{r_q}}\frac{1}{2\pi i} \oint_{\a_1} \frac{(z-m)^{{r_q}-1}dz}{y}=\frac{g_{N_c-{r_q}}}{{r_q} \sqrt{{W\p}^2+f}|_{z=m}} \ .
\eeq
thus we obtain
\beq
\label{res}
\T = \left.\sqrt{{W\p}^2+f}\right|_{z=m} \ .
\eeq 
The dual-quark condensate is
\bea
\tD_I D_J=
\left(
\begin{array}{cccc}
-\left.\sqrt{{W\p}^2+f}\right|_{z=m} {\bf 1}_{{\rm min}(r,N_f-r)}&&&\\
&0 &&\\
&& \ddots &\\
&&& 0 \\
\end{array}\right)\ .
\eea

\section{The chiral ring \label{chiral}}

The generators of the chiral ring of the theory are:
\beq
\label{convenction}
T(z)=\Tr\, \frac{1}{z-\Phi}\ ,\qquad
R(z)=-\frac{1}{16\sqrt{2}\pi^2}\Tr\,\frac{W^{\a}W_{\a}}{z-\Phi} \ ,
\eeq
\beq
M_{IJ}(z)=\tQ_I \frac{1}{z-\Phi}Q_J \ .
\eeq
The generalized Konishi anomalies provides a solution for the chiral ring \cite{CDSW}. The anomalies  that we need are the following:
\bea
\label{kenanomaly}
&&[W\p(z)R(z)]_-= R(z)^2 \ ,\\
&&[M_{II}(z) (z-m)]_-= R(z) \ , \\
&&[M_{IJ}(z) (z-m)]_-=0 \qquad I \neq J \ .\nonumber
\eea 
The solution of the first equation is
\beq
\label{R}
2R(z)=W\p(z)-\sqrt{W\p(z)^2+f(z)} \ ,
\eeq
where $f(z)$ is a polynomial of degree $k-1$ that depends on the vacuum. By (\ref{R}) we are naturally led to consider the Riemann surface $\Sigma_{\N=1}$ defined by the equation
\beq
\label{SigmaN=1}
{y_m}^2=W\p(z)^2+f(z) \ .
\eeq
This is a double sheeted cover of the complex plane on which $R(z)$ is uniquely defined. We call $q$ and $\tq$  the two points of $\Sigma_{\N=1}$ with the same  coordinate $z=m$.  
For every flavor $I$ we can choose if it is locked ($r_I=1$) or not ($r_I=0$) and the vacua will be characterized by $r=\sum_{I=1}^{N_f} r_I$.   When $r_I=0$  $M_{II}(z)$  must be regular in $q$
\beq
M_{II}(z)=\frac{R(z)}{z-m}-\frac{R(q)}{z-m} \ .
\eeq
When $r_{I}=1$ one can find the solution by continuously deforming the theory, in such a way that the pole passes from the second to the first sheet. The solution is:
\beq
\label{M}
M_{II}(z)=\frac{R(z)}{z-m}-\frac{W\p(m)-R(q)}{z-m} \ .
\eeq
We take the off diagonal terms $M_{IJ}=0$ since we are at the root of the Higgs branch.

In the case of degenerate masses is possible to express the dual-quark condensate (\ref{res}) as a simple quantity in the chiral ring of the theory.
 We have to distinguish two cases. 
If $r \leq N_f/2$, the condensate $ \tD_{I} D_{J} $ is the residue of the differential $M_{IJ}(z) dz$ around the point $z=m$ in the first sheet
\beq
\label{aafirst}
\tD_{I} D_{J}= \frac{1}{2\pi i} \oint_{q} M_{IJ}(z) dz  \ .
\eeq
If $r \geq N_f/2$  the condensate $ \tD_{I} D_{J} $ is the residue of the differential $M_{IJ}(z) dz$ around the point $z=m$ in the second sheet
\beq
\label{afirst}
\tD_{I} D_{J}= \frac{1}{2\pi i} \oint_{\tq} M_{IJ}(z) dz   \ .
\eeq

\section{ Conclusion  \label{conclusion}      }

The vortices considered here are exactly BPS only when the superpotential is a linear function of $\Phi$.   As studied in \cite{Stefano} they have non-BPS corrections and the  condition under which they are small with respect to the BPS tension is
\beq
\label{g}
\frac{ g(\mu)^2 \,  {{W_{eff}\p}\p}^2}{W_{eff}\p} << 1\ ,
\eeq
where we have considered $W_{eff}$ that enters in (\ref{lowsup}). The energy scale $\mu$ of the $U(1)$ breaking is roughly $\sqrt{W_{eff}\p}$. We argue that a region of parameters exists  where the condition (\ref{g}) is satisfied. To find it we multiply the tree level superpotential by a constant $\epsilon$:
\beq
\epsilon W(z)\ , \qquad 0\leq \epsilon \leq 1 \ .
\eeq 
If we send $\epsilon \to 0$, the BPS tension goes to zero like $ \epsilon$ while the non-BPS correction goes to zero more quickly. In fact ${{W_{eff}\p}\p}^2$ brings a factor $\epsilon^2$ and $e^2(\mu)$ vanishes logarithmically with $\epsilon$. Thus for sufficient little $\epsilon$ our vortices are almost BPS.\footnote{If ${{W_{eff}\p}\p}^2/W_{eff}\p \leq 1$, the strong coupling region, where $e^2(\mu)$ is small, is enough for (\ref{g}) to be valid.}

The new result of this paper is the expression of the dual-quark condensate as a simple quantity in the chiral ring of the theory. The dual-quark has no meaning out of the strong coupling regime where the dual $SU(r_q) \times U(1)$ is weakly coupled. The chiral ring quantity in the right hand side of  (\ref{aafirst}) and (\ref{afirst}) is instead computed using the generalized Konishi anomaly and thus is valid in every regime, not only at strong coupling. Unfortunately out of the strong coupling we have no more control of the non-BPS corrections.

Equations (\ref{aafirst}) and (\ref{afirst}) are a signal of the duality $r \leftrightarrow N_f-r$. The vacuum counting of \cite{CKM} and more recently \cite{Marm}, shows that classical vacua with $r$ or $N_f-r$ locked quarks have the same low energy description in terms of $r_q$-vacua. Thus we see that the duality $r \leftrightarrow N_f-r$ appears in equations (\ref{aafirst}) and (\ref{afirst}) as an exchange between the first and the second sheet of the $\N=1$ Riemann surface.

\appendix

\section{Nonabelian Vortex  \label{NAV}}

Now we recall in brief the construction of the nonabelian BPS vortex.  The Lagrangian is $SU(n_c) \times U(1)$ (in general with different couplings), plus a charged scalar field in the $\underline{r},1$ and the BPS potential
\beq
{\cal L} = -\frac{1}{4{g_{na}}^2} (F_{\mu\nu}^a)^2 -\frac{1}{4g^2} (F_{\mu\nu})^2 - (D_\mu q)(D^\mu q)-\frac{{g_{na}}^2}{2}\sum_a(q^{\dagger } T^a q)^2-
\frac{g^2}{2}(q^{\dagger }q-2 n_c v)^2\ .
\eeq
When $g_{na}=g$ the potential is equal to (\ref{potential}).
\beq
D_{\mu}q=\partial_{\mu}q-iA_{\mu}^a T^a q-iA_{\mu}q
\eeq
The potential is minimized when
\beq
\langle q \rangle=\left(\begin{array}{ccccc}
\sqrt{2v}&&&0&\\
&\ddots&&&\ddots\\
&&\sqrt{2v}&&\\
\end{array}\right) \ .
\eeq
This VEV breaks the flavor symmetry of the Lagrangian
\beq
\label{ressimm}
U(n_f) \rightarrow U(n_c)_{C+F} \times U(n_f-n_c) \ .
\eeq
The BPS tension is a boundary term
\beq
T_{BPS}=\int d^2 x \; \de_k\epsilon_{kl}  \left( 2 A_l v -iq^{\dagger }D_lq \right)
\eeq
To build the vortex configuration we embed the ordinary $U(1)$ vortex in this theory.  All such embeddings are $U(r)$ rotations of
\bea
\label{vortexconf}
q&=&\sqrt{2v} \left(\begin{array}{cccccc}
e^{i\theta}q_1(r)&&&&0&\\
&q_2(r)&&&&\ddots\\
&&\ddots&&&\\
&&&q_2(r)&&\\
\end{array}\right)\ ,\\
{A_{na}}_{\, k}&=&-\epsilon_{kl}\frac{\hat{r}_l}{r}f_{na}(r)\left(\begin{array}{cccc}
(n_c-1) &&&\\
&-1&&\\
&&\ddots&\\
&&&-1\\
\end{array}\right) \ , \nonumber \\
{A}_k&=&-\epsilon_{kl}\frac{\hat{r}_l}{r}f(r) \ .
\nonumber
\eea
where $q_1(r)$, $q_2(r)$, $f(r)$ and $f_{na}(r)$ are some profile functions that satisfy the boundary conditions $q_1(0)=f(0)=f_{na}(0)=0$,  $q_1(\infty)=q_2(\infty)=1$ and $f(\infty)=f_{na}(\infty)=1/n_c$.
The $r$ independent vortices constructed this way are degenerate with tension
\beq
T_{BPS}=4\pi{v} \ .
\eeq
The vortex solution (\ref{vortexconf}) classically breaks the residual global symmetry (\ref{ressimm}) to $U(1) \times U(n_c-1)$. This leads to the existence of a moduli space. 
\beq
q \rightarrow U q U^{-1} \ , \qquad A_{na} \rightarrow U A_{na} U^{-1} \ .
\eeq
the moduli space is:
\beq
\mathbb{CP}^{n_c-1}=\frac{ U(n_c)_{C+F} }{ U(1) \times U(n_c-1)} \ .
\eeq
When $n_f>n_c$ other zero modes are present in our theory.
Hanany and Tong
\cite{HT} have found that among the classical solutions for $n_f>n_c$ are semi-local vortices.

\section* {Acknowledgments}

 I would like to  thank for useful discussions Sergio Benvenuti, Kenichi Konishi, Giacomo Marmorini and  Marco Matone.


\begin{thebibliography}{}

\bibitem{Stefano}
S.~Bolognesi,
{\it The Holomorphic Tension of Vortices}, JHEP {\bf 0501} (2005) 044,
hep-th/0411075.



\bibitem{APS}
P. C.~Argyres, M. R.~Plesser and N.~Seiberg,
{\it The Moduli Space of N=2 SUSY QCD and Duality in N=1 SUSY QCD},
Nucl.\  Phys.\  B\ {\bf 471}  (1996) 159,
hep-th/9603042.

\bibitem{CKM}
G.~Carlino, K.~Konishi and H.~Murayama,
{\it Dynamical Symmetry Breaking in Supersymmetric $SU(N_c)$ and $USp(2N_c)$ Gauge Theories},
Nucl.\  Phys.\ B\  {\bf 590}  (2000) 37,
hep-th/0005076.


\bibitem{vortici}
R.~Auzzi, S.~Bolognesi, J.~Evslin, K.~Konishi and A.~Yung,
{\it Nonabelian superconductors: vortices and confinement in N = 2 SQCD},
Nucl.\ Phys.\ B {\bf 673} (2003) 187,
hep-th/0307287.


\bibitem{HT}   A.~Hanany, D.~Tong,
{\it Vortices, instantons and branes},
hep-th/0306150; 
{\it Vortex strings and four-dimensional gauge dynamics},
JHEP {\bf 0404} (2004) 066,
hep-th/0403158.








\bibitem{largeN}
F.~Cachazo, K.~A.~Intriligator and C.~Vafa,
{\it A large N duality via a geometric transition},
  Nucl.\ Phys.\ {\bf B603} (2001) 3-41,
hep-th/0103067.



\bibitem{CDSW}
F.~Cachazo, M.~R.~Douglas, N.~Seiberg and E.~Witten,
{\it Chiral rings and anomalies in supersymmetric gauge theory},
  JHEP\ {\bf 0212}  (2002) 071,
hep-th/0211170;

F.~Cachazo, N.~Seiberg and E.~Witten,
{\it  Phases of N=1 supersymmetric gauge theories and matrices},
 JHEP\ {\bf 0302} (2003) 042,
hep-th/0301006.





\bibitem{factor}

Y.~Ookouchi,
{ \it N = 1 gauge theory with flavor from fluxes},
JHEP {\bf 0401} 014 (2004),
hep-th/0211287.

V.~Balasubramanian, B.~Feng, M.~x.~Huang and A.~Naqvi,
{\it Phases of N = 1 supersymmetric gauge theories with flavors},
Annals Phys.\  {\bf 310} 375 (2004), hep-th/0303065.


\bibitem{CSW}

F.~Cachazo, N.~Seiberg and E.~Witten,
{\it Chiral rings and phases of supersymmetric gauge theories},
 JHEP\ {\bf 0304} (2003) 018,
hep-th/0303207.


\bibitem{Marm}
S.~Bolognesi, K.~Konishi and G.~Marmorini,
{\it Light nonabelian monopoles and generalized r-vacua in supersymmetric gauge theories}, hep-th/0502004.





%\bibitem{RioandJarah}
%R.~Auzzi, S.~Bolognesi and J.~Evslin,
%{\it Monopoles can be confined by 0,1 and 2 vortices},
%hep-th/0411074.
















\end{thebibliography}
\end{document}